\documentclass[pra,onecolumn,preprintnumbers,superscriptaddress]{revtex4} 

\usepackage{graphicx}
\usepackage{bm}
\usepackage{amsmath}
\usepackage{amssymb}
\usepackage{amsfonts}
\usepackage{fancyhdr}
\usepackage{slashed}
\usepackage{latexsym,epsfig,bbm}

\def\>{\rangle}
\def\<{\langle}
\def\be{\begin{equation}}
\def\ee{\end{equation}}
\def\ft{\footnotesize}

\begin{document}

\title{Arbitrary-dimensional teleportation of optical number states with linear optics}

\author{T. Farrow}
\email{tristan.farrow@physics.ox.ac.uk}
\affiliation{Atomic and Laser Physics, Clarendon Laboratory, University of Oxford, Parks Road, Oxford, OX1 3PU, United Kingdom}
\affiliation{Centre for Quantum Technologies, National University of Singapore, 3 Science Drive 2, 117543, Singapore}
\author{V. Vedral}
\affiliation{Atomic and Laser Physics, Clarendon Laboratory, University of Oxford, Parks Road, Oxford, OX1 3PU, United Kingdom}
\affiliation{Centre for Quantum Technologies, National University of Singapore, 3 Science Drive 2, 117543, Singapore}

\date{\today}

\begin{abstract}
Quantum state teleportation of optical number states is conspicuously absent from the list of experimental milestones achieved to date~\cite{braun15}. Here we demonstrate analytically a teleportation scheme with fidelity $100\%$ for optical number states of arbitrary dimension using linear optical elements only. To this end, we develop an EPR source to supply Bell-type states for the teleportation, and show how the same set-up can also be used as a Bell-state analyser (BSA) when implemented in a time-reversal manner. These two aspects are then brought together to complete the teleportation protocol in a scheme that can deliver perfect fidelity, albeit with an efficiency that decays exponentially as the occupation of the number states increases stepwise. The EPR source and BSA schemes both consist of two optical axes in a symmetrical V-shape experimental layout, along which beam-splitters are placed cross-beam fashion at regular intervals, with their transmittivities treated as variables for which the values are calculated \emph{ad hoc}. In particular, we show the full treatment for the case of qutrit teleportation, and calculate the transmittivity values of the beam splitters required for teleporting qubits, qutrits, qupentits, quheptits and qunits. The general case for arbitrary-dimensional number state teleportation is demonstrated through a counting argument.
\end{abstract}
\maketitle 

Photons (interchangeably referred to as optical number-states or
Fock-states) are bosons. Unlike fermionic electrons, they have the property
that they can fly undisturbed since they do not interact with each other. This can be used to great advantage particularly in
areas such as quantum communication and information processing where
an ideal transmission medium would have the property that it could
carry information without degrading it, i.e. that the received
information would have a perfect fidelity of $1$. In this respect, photons
represent the medium of choice and offer significant advantages over
electronic channels. But since they do not interact, they are
also difficult to control (compared to electrons).
\\\\This paper offers a generalisation of the teleportation
protocol to an arbitrary dimension of optical number states. Its
generality sets it apart from previous teleportation schemes. The
first laboratory implementations of teleportation, performed by
Bouwmeester et al.~\cite{bouw} and Boschi et al.~\cite{DM2},
succeeded in teleporting a photon's polarisation state, that is, an
internal degree of freedom. A more general scheme was implemented by Furusawa et al.~\cite{caltech}, where a coherent state of light was teleported via continuous variables~\cite{braunstein}, that is, a subset of the
spatial degree of freedom. A third and more general
type of teleportation was realised for a single optical number state
in superposition with the vacuum state~\cite{dolmio}. We take this
as our starting point, and extend the idea of number state
teleportation to arbitrarily large states.


\section*{Introducing teleportation}\label{secn introConcepts}
Since teleportation was first introduced as a concept in 1993
\cite{bennnett} it has also been implemented in trapped ions~(\cite{bar04} -\cite{noe}), cold atomic ensembles~(\cite{sher} -\cite{bao}), solid state and NMR systems~(\cite{gao} - \cite{niel}), as well as in optical-modes~(\cite{caltech} -\cite{tak}), photonic polarisation~(\cite{bouw}, \cite{ur}-\cite{ma12}) and spin-orbital angular momenta states~(\cite{wang15}). 
It offers a fundamentally new method of transferring
quantum information between two particles separated by an arbitrary
distance, currently over 140km in free space~\cite{zeil12}. For the scheme to work, the two particles must be
in a non-local superposition of quantum states, otherwise known as
entanglement, a notion that came with the Einstein-Rosen-Podolsky (EPR)
paradox~\cite{EPR}. Within the context of a teleportation scheme,
two entangled particles are said to form a \emph{quantum channel},
as opposed to a classical one. A corollary is the fact that all
operations performed in a teleportation protocol are \emph{local},
so that no measurements need be carried out on all the states in the
system simultaneously.\label{local ops}

\begin{figure}
\includegraphics[width=15cm]{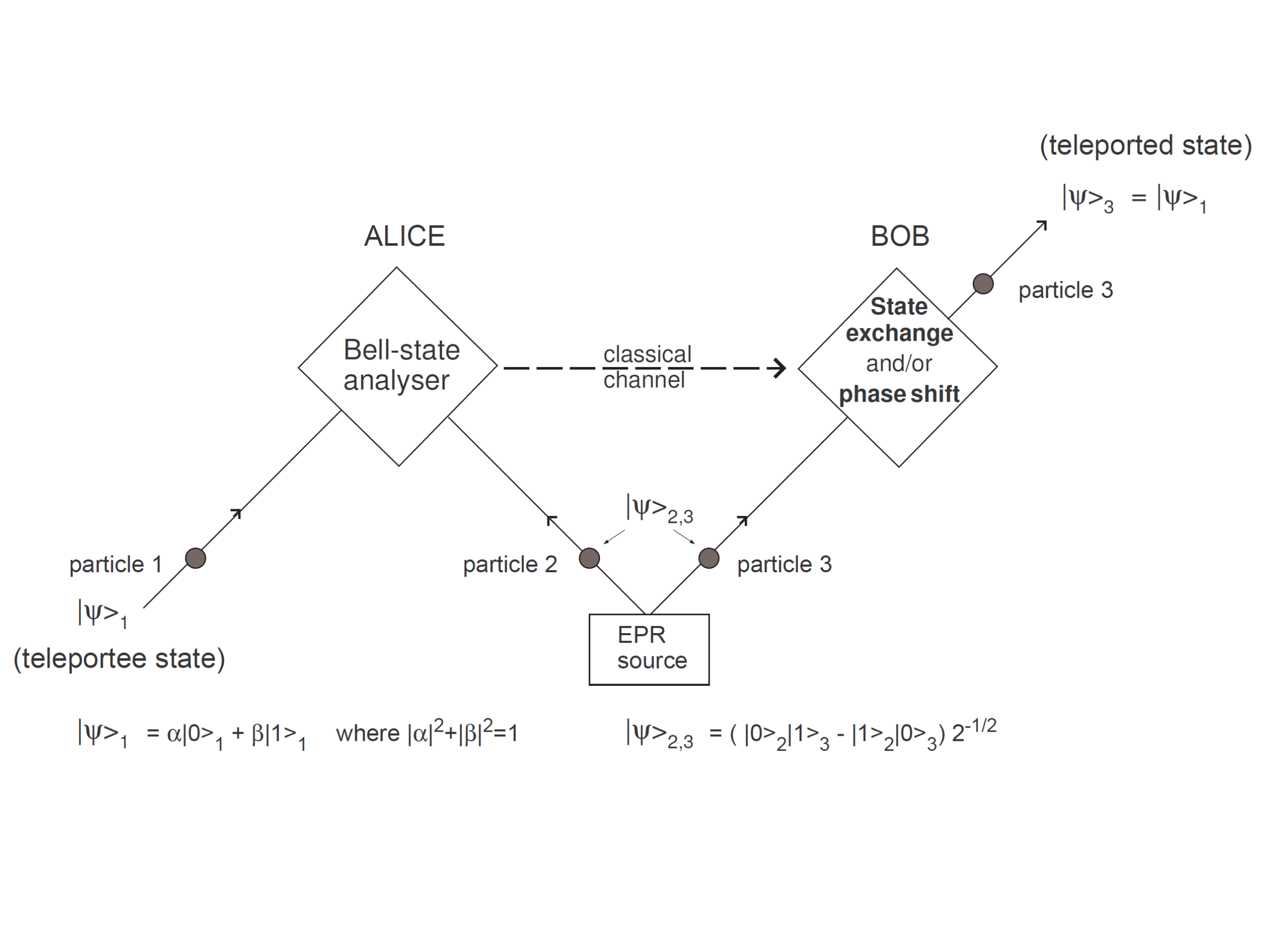}
\caption{Protocol for teleporting a pure quantum state from particle 1, i.e. a linear superposition of qubits between logical $0$ and $1$ describing the state of the particle, to particle 3 some distance away. Particle 3 is generated by an EPR source jointly with particle 2 in a maximally entangled singlet state described by $|\Psi\>_{2,3}$, which establishes a quantum communication channel. The result of the Bell-state Analysis~\cite{phil05} performed by Alice is communicated to Bob via a classical channel, who then recovers the state on particle 3, thus completing the teleportation. The initial state on particle 1 is destroyed (maximally mixed).}
\label{genericTelProtocol}
\end{figure}

A fundamental distinction between teleportation and classical communication stems from the \emph{no-cloning theorem}~\cite{wooters} in quantum mechanics, which states that no \emph{copy} of a quantum state can be made. Note that this
does not mean a quantum state cannot be \emph{transferred} from one
particle to another, but merely that a reproduction of it cannot
coexist with the original. In practise this means that during a
teleportation, the initial quantum state is destroyed, by becoming maximally mixed, so that at the end we are not left with any
copies of the state with which we started. Rather, only the
teleported state remains, and this will have been transferred across
to the recipient particle where it will now reside, as illustrated in figure~\ref{genericTelProtocol}. Unlike classical
communication, teleportation is not a facsimile process.

\section*{Generating maximally entangled states with linear optics: EPR
source}
\label{secn EPRsource}

\begin{figure}
\includegraphics[width=13cm]{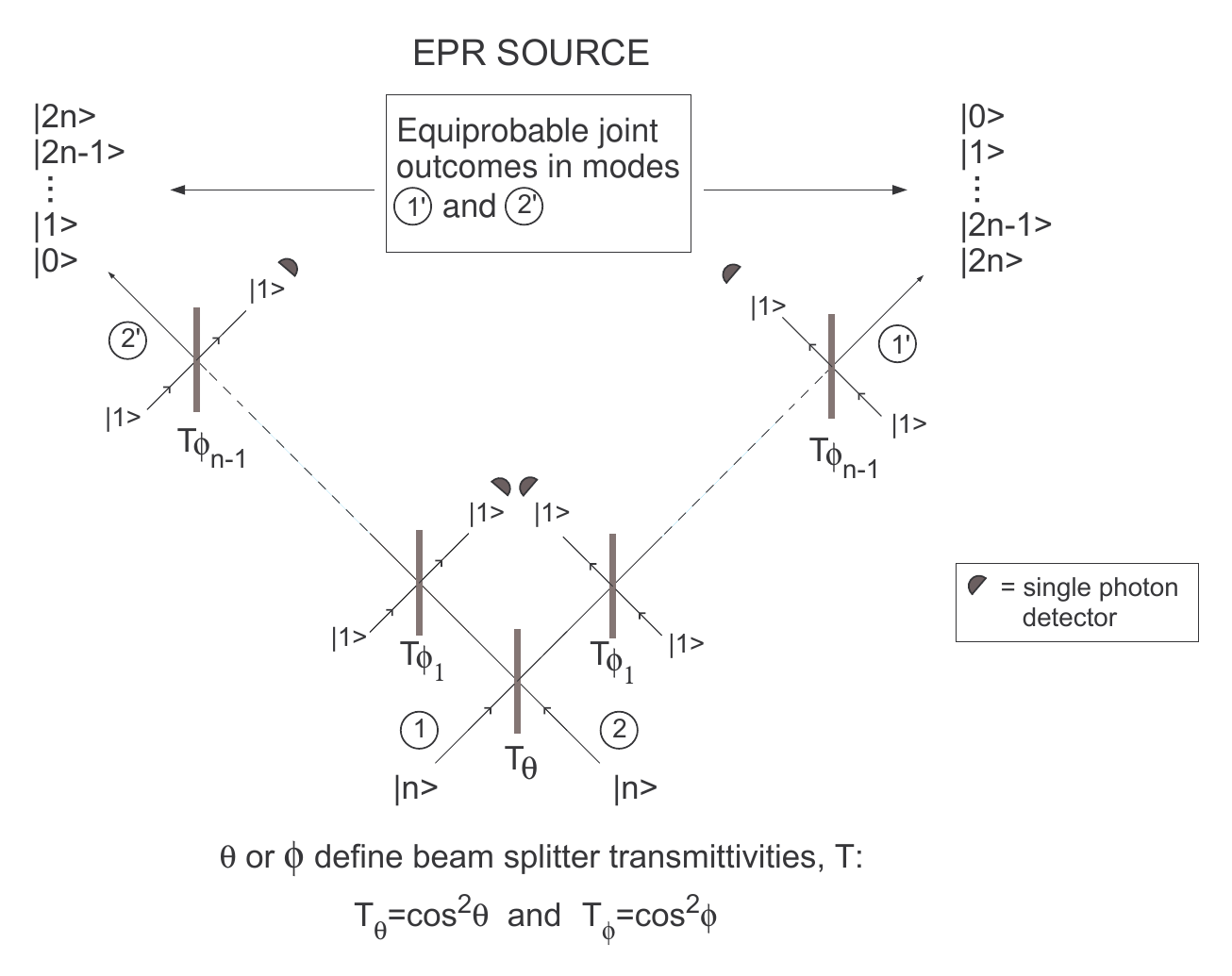}
\caption{Generic schematic of our proposed EPR source. A
symmetric photon input is injected into channels $1$ and $2$, and we
sub-select \emph{only} the outcomes (i.e. the set of states and
their amplitudes) that result in single photons emerging from all
outputs \emph{except} the two outermost (labelled $1'$ and $2'$). An
implementation of the scheme would have to rely on single-photon
detection schemes.}
\label{epr(nn)}
\end{figure}

To generate maximally entangled states with linear optics~\cite{laflamme}, we
propose a set-up consisting of beam splitters placed along two
branches in a symmetrical ``V'' layout, as shown in
figure~\ref{epr(nn)}. Each branch contains an equal number of
beam-splitters while their transmittivities \emph{along} each branch
are treated as variables (from $T_\theta$ for the first
beam-splitter, up to $T_{\phi_{n-1}}$ for the last). Beam-splitters
at the same distance along the two branches have the same
transmittivity. An equal number of
photons is injected into the inputs of the two branches (labelled
$1$ and $2$), and \emph{only} the outcomes that result in single
photons emerging from all outputs \emph{except} the two outermost
(labelled $1'$ and $2'$) are selected. Provided each of the
detectors along the two branches registers a single photon, then we
can be sure that the possible combination of output states in modes
$1'$ and $2'$ will range from \be
|2n\>_{1'}|0\>_{2'},|2n-1\>_{1'}|1\>_{2'},\ldots
,|1\>_{1'}|2n-1\>_{2'},|0\>_{1'}|2n\>_{2'} \label{eq
sub-outcomes}\ee where $|n\>$ is the number state injected into each
of the two branches. A laboratory implementation of the scheme would
require single-photon detection devices to be located along all the
side-outputs of the two branches, as shown in figure~\ref{epr(nn)}. Apart from the qubit, we consider only even number inputs
of photons, although the scheme can be extended to assymmetric
inputs too.
\newline Each outcome, or state, in equation (\ref{eq
sub-outcomes}) will have associated with it a probability amplitude as
a function of the transmittivities of the beam-splitter in the two
branches: \be
|State\>_{out}=A|2n\>_{1'}|0\>_{2'}+B|2n-1\>_{1'}|1\>_{2'}+\ldots+C|n\>_{1'}|n\>_{2'}+
D|1\>_{1'}|2n-1\>_{2'}+E|0\>_{1'}|2n\>_{2'} \label{eq
sumsuboutcomes} \ee where $A,B,C,D,E$ represent expressions for the
amplitudes in terms of variables $T_\theta$ to $T_{\phi_{n-1}}$. We
note that since they represent the probability amplitudes of a
sub-selection of possible outcomes, we normalise them so that
$|A|^2+|B|^2+|C|^2+|D|^2+|E|^2=1$. For example, let us imagine that
we have five expressions for $A$ to $E$, then a set of two
independent simultaneous equations can be formed ($A=B$ and $B=C$),
which can then be solved for the transmittivity values. At these
values, the probability amplitudes are equal, since $A=B=C$, while
by symmetry $D=B$ and $E=A$. The resulting sum of the state vectors
is then our maximally entangled state. Note that we only need a
simultaneous solution to the first three expressions ($A,B,C$) due
to the symmetry of our system, which ensures automatically that
$B=D$ and $A=E$, meaning that symmetrical outcomes, say,
$|2n\>_{1'}|0\>_{2'}$ and $|0\>_{1'}|2n\>_{2'}$, are equally likely.
\\The total number of beam splitters
in each branch (including the first beam-splitter shared by the two
branches) equals the number of photons injected into each branch. If
a symmetrical input of $|n\>_1$ photons is injected into branch $1$
and $|n\>_2$ into branch $2$, then the total number of beam-spitters
will be $2n-1$, including the shared first element.\\
\\A general counting argument follows, together with a method
 for calculating the transmittivities that produce maximally entangled
 states. Larger input states rapidly become too onerous for analytical treatment,
 so warrant numerical simulation.

\subsection*{Input $|1\rangle_1|1\rangle_2$}\label{secn input11}

\begin{figure}
\includegraphics[width=6cm]{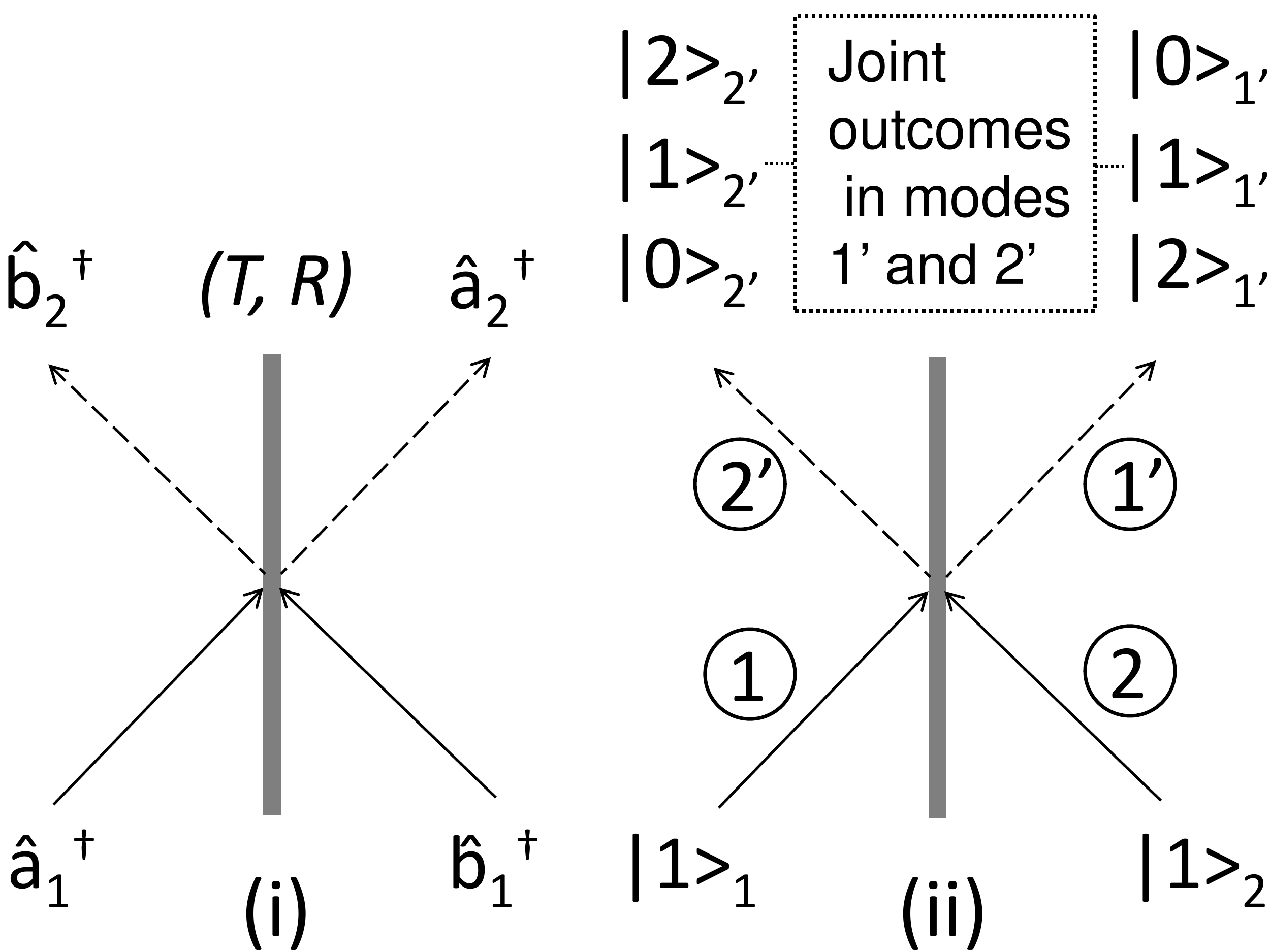}
\caption{(i) Beam splitter with transmittivity $T$ and reflectivity $R$, where $T+R=1$ since $T=cos^{2}\theta$ and $R=sin^{2}\theta$. In the operator model, the unitary transformation due to a beam splitter on an optical mode can be expressed algebraically as
$\hat{a_1}^\dagger\stackrel{bs}{\longmapsto}
\cos\theta\hat{a_2}^\dagger+\sin\theta\hat{b_2}^\dagger$ \,and\,
$\hat{b_1}^\dagger\stackrel{bs}{\longmapsto}-\sin\theta\hat{a_2}^\dagger+\cos\theta\hat{b_2}^\dagger$,
where the input and output modes $\hat{a}^\dagger$ and $\hat{b}^\dagger$ are shown in the figure. This can be represented as a matrix to highlight the beam splitter's mapping function and unitarity. (ii) EPR source: using an asymmetric beam-splitter the outcome $|11\>_{1',2'}$ can be achieved with equal probability as $|20\>_{1',2'}$ and $|02\>_{1',2'}$.
}
\label{bs}
\end{figure}

For an input consisting of one photon on each input mode, we have \[
|1\rangle|1\rangle\equiv\hat{a}^\dagger\,\hat{b}^\dagger\,
|0\rangle|0\rangle
\]
\[
\stackrel{bs}{\longmapsto}\,\big(\sqrt{T}\,\hat{c}^\dagger+\sqrt{R}\,\hat{d}^\dagger\big)\,\big(-\sqrt{R}\,\hat{c}^\dagger+\sqrt{T}\,\hat{d}^\dagger\big)\,|0\rangle|0\rangle
\]
\[
=\,\Big(
T\,\hat{a}^\dagger\hat{b}^\dagger-\sqrt{TR}(\hat{a}^\dagger)^2+\sqrt{TR}(\hat{b}^\dagger)^2-R\hat{b}^\dagger\hat{a}^\dagger\Big)\,|0\rangle|0\rangle
\]
\[
=\,\Big(
(T-R)\,\hat{a}^\dagger\hat{b}^\dagger+\sqrt{TR}\big(\hat{b^\dagger}^2-\hat{a^\dagger}^2\big)\Big)\,|0\rangle|0\rangle
\]
\be \therefore\,\,|1\rangle_1|1\rangle_2\,\,\,
 \stackrel{bs}{\longmapsto}
(T-R)\,|1\rangle_{1'}|1\rangle_{2'}+\sqrt{TR}\,\sqrt{2}\,\big(|0\rangle_{1'}|2\rangle_{2'}-|2\rangle_{1'}|0\rangle_{2'}\big)
\label{eq prefactors11} \ee

For $T=R$, the output $|1\rangle_{1'}|1\rangle_{2'}$ does not arise,
while $|0\rangle_{1'}|2\rangle_{2'}$ and
$|2\rangle_{1'}|0\rangle_{2'}$ are equiprobable. In this scenario,
the resulting superposition is not a Bell-type state. For maximum
entanglement, we require that all the amplitude prefactors in
equation~(\ref{eq prefactors11}) be equal. Hence we form the
simultaneous equations~(\ref{eq epr11 1}) and (\ref{eq epr11 2})
which we solve for $T$ and $R$.  The transmittivity value, $T$,
which satisfies both equations causes all the three states
($|1\rangle|1\rangle, |0\rangle|2\rangle$ and $|2\rangle|0\rangle$)
to have an equal amplitude, and so their superposition to constitute
a Bell-type state.

\begin{equation}
(T-R)=\sqrt{2}\sqrt{TR} \label{eq epr11 1}
\end{equation}

\begin{equation}
T+R=1 \label{eq epr11 2}
\end{equation}

Solving the two equations yields:
\begin{equation}
T=0.211325
\end{equation}
\begin{equation}
R=0.788675
\end{equation}
\begin{equation}
\mbox{Efficiency: }0.57735\times\frac{1}{3^2}
\end{equation}

Employing a beam splitter of this transmittivity ensures that $60\%$
of the output from our EPR set-up is a maximally entangled state of
the form
\[
\frac{1}{\sqrt{3}}\big(|20\>+|11\>+|02\> \big).
\]
This Bell-type state can then be used in a teleportation scheme for
qutrits ($\alpha|0\>+\beta|1\>+\gamma|2\>$). The normalisation factor
$\frac{1}{9}$ comes from the fact that for a qutrit, there are nine
possible Bell-states (since $d=3$ and $d^2$ is the number of
Bell-states), from which we can detect only one.

\subsection*{Input $|2\rangle_1|2\rangle_2$}\label{secn input22}

\begin{figure}
\includegraphics[width=10cm]{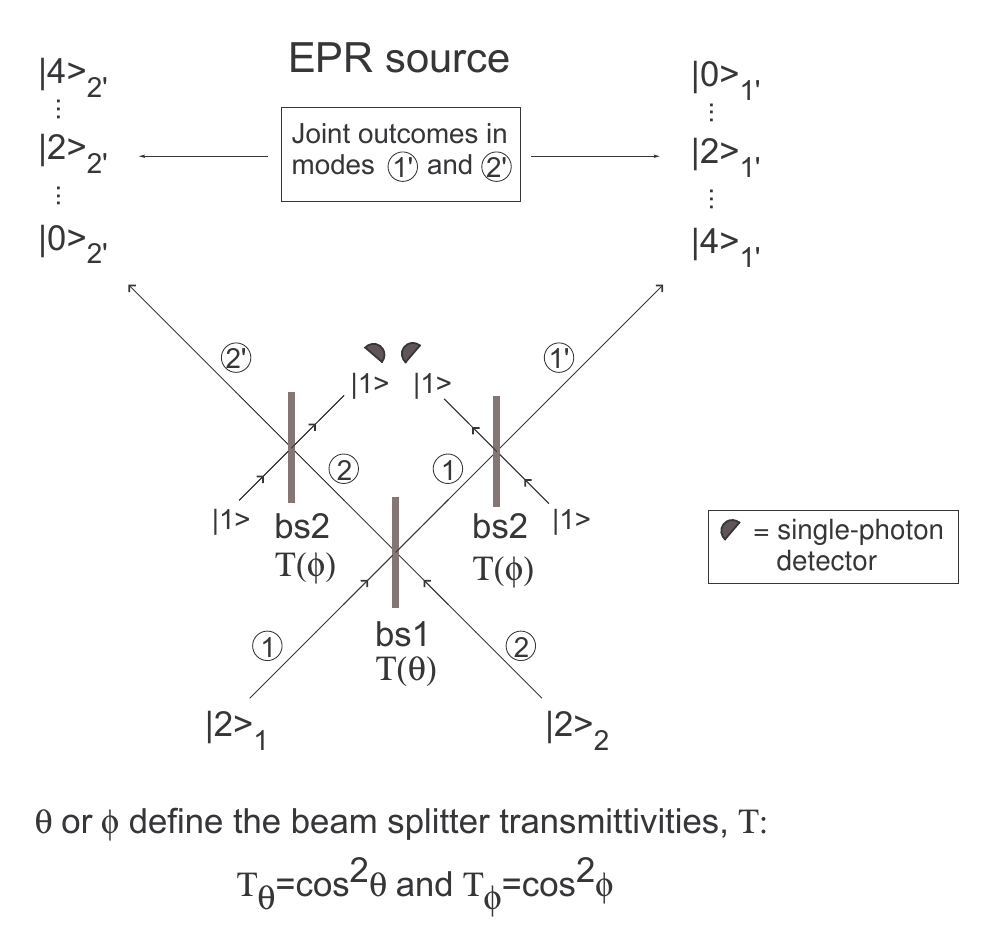}
\caption{Encircled numbers $1'$ and $2'$ label specific
modes.}
\label{epr(22)}
\end{figure}

In the case of the input  $|2\rangle_1|2\rangle_2$, we observe an
exponential increase in the number of terms in the amplitude
prefactors, albeit the number of state vectors increases only
polynomially compared to the $|1\rangle_1|1\rangle_2$ input. This is
a trend we observe for every consecutive increase in the number of
photons. For this reason, we do not show here the complete
calculations for the beam-splitter transmittivites, but rather limit
ourselves to illustrating the exponential proliferation of terms in
the amplitudes. We then give the solutions for the transmittivites
required to generate a Bell-state of the form shown in
equation~\ref{eq bell epr22}. The values were computed numerically
and can in principle be performed for any $n$.
\\An input of $|2\rangle_1|2\rangle_2$ will produce the following
Bell-type state: \be
A|4\>_{1'}|0\>_{2'}+B|3\>_{1'}|1\>_{2'}+C|2\>_{1'}|2\>_{2'}+D|1\>_{1'}|3\>_{2'}+E|0\>_{1'}|4\>_{2'}
\label{eq bell epr22}\ee

where $A=B=C$. By symmetry, $D$ and $E$ are equal to $A$ and $B$
respectively.

Proceeding as in the case for the $|1\>_1|1\>_2$ input, we express
the $|2\>_1|2\>_2$ in terms of creation operators acting on the
vacuum state:

\[
|2\rangle_1|2\rangle_2\equiv\frac{\hat{a}^{\dagger^2}}{\sqrt{2}}\,\frac{\hat{b}^{\dagger^2}}{\sqrt{2}}\,
|0\>_1|0\>_2
\]where $\hat{a}^{\dagger}$ and $\hat{b}^{\dagger}$ act on modes $1$ and $2$ respectively.
The action of beam-splitter $bs1$ is expressed using the convention outlined in figure~\ref{bs} and leads to the expansion
\[
\stackrel{bs1}{\longmapsto}\frac{1}{2}\,\big(\sqrt{T}\,\hat{a}^\dagger+\sqrt{R}\,\hat{b}^\dagger\big)^2\,\big(-\sqrt{R}\,\hat{a}^\dagger+\sqrt{T}\,\hat{b}^\dagger\big)^2\,|0\rangle|0\rangle
\]
Following the action of $bs1$, the overall state is incident on a
second beam-splitter, $bs2$, where a second transformation is
applied:

\[
\stackrel{bs2}{\longmapsto}\frac{1}{2}\,\Bigg[
\Bigg(\sqrt{T}\,\Big(\sqrt{T}\,\hat{a}^\dagger+\sqrt{R}\,\hat{b}^\dagger\Big)+\sqrt{R}\,\Big(-\sqrt{R}\,\hat{a}^\dagger+\sqrt{T}\,\hat{b}^\dagger\Big)\Bigg)^2
\]
\[
+\Bigg(-\sqrt{R}\,\Big(\sqrt{T}\,\hat{a}^\dagger+\sqrt{R}\,\hat{b}^\dagger\Big)+\sqrt{T}\,\Big(-\sqrt{R}\,\hat{a}^\dagger+\sqrt{T}\,\hat{b}^\dagger\Big)\Bigg)^2\,\Bigg]|0\rangle|0\rangle
\]

Expanding the expression leads to exponents of $4$ in
$\hat{a}^\dagger$ and $\hat{b}^\dagger$ in addition to cross-terms.
It can readily be noticed that the number of terms in the expression
has increased exponentially on the preceding simple input
$|1\>_1|1\>_2$.
\\\\In the set-up shown in figure~\ref{epr(22)}, we have only two different transmittivities,
$T_\theta$ and $T_\phi$. Their values are calculated by solving
simultaneously the expressions for the amplitude prefactors $A,B,C$
for the output states
\[
A|4\>_{1'}|0\>_{2'},B|3\>_{1'}|1\>_{2'},C|2\>_{1'}|2\>_{2'}
\]
obtained from equation~\ref{eq bell epr22}. Although it might at
first appear that we have three equations and only two variables,
$A,B,C$ are in fact only three expressions, which we set to equal
one another. We solve them in pairs so that (\ref{a=b}) and
(\ref{b=c}) now constitute a properly formed set of two simultaneous
equations. The condition that $C=A$ follows.

\begin{equation} A=B \label{a=b}
\end{equation}
and \begin{equation} B=C \label{b=c}
\end{equation}

We solved the equations numerically, expressing them in terms of
trigonometric functions with the advantage that the condition
$T+R=1$ is satisfied automatically.
\[
A=-\sqrt{6}\cos^4\phi\,\cos^2\theta\,\left(-1+\cos^2\theta\right)\left(-4+5\cos\phi^2\right)
\]
\[
B=\sqrt{6}\cos^2\phi\,\sin\theta\,\cos\theta
\big(-3+10\cos^2\phi-8\cos^4\phi\,
\]
\[
\,\,\,\,\,\,\,\,\,\,\,\,\,\,\,\,\,\,\,\,\,\,\,\,\,\,\,\,\,\,\,\,\,\,\,\,\,\,\,\,\,\,\,\,\,\,\,\,\,\,\,\,\,\,\,
\,\,\,\,\,\,\,\,\,\,\,\,\,\,\,\,\,\,\,\,\,\,\,\,\,\,\,\,\,+\,6\cos^2\theta-20\cos^2\phi\,\cos^2\theta+16\cos^4\phi\,\cos^2\theta\big)
\]
\[
\,\,\,C=\cos^2\phi\,\left(1-6\cos^2\theta + 6
cos^4\theta\right)\left(2\sin^2\phi-\cos^2\phi\right)^2
\]

where $\cos^2\theta=T_{\theta},\, \sin^2{\theta}=R_{\theta}, \,
\cos^2\phi=T_{\phi}$ \,and\, $R_{\phi}=\sin^2\phi=R_{\phi}$.
\\Solving these expressions for $\theta$ and $\phi$ (i.e. $T_{\theta},\,T_\phi$) for a symmetrical input
of $|2\>_1|2\>_2$ gives the following transmittivity values
(reflectivity is $R=1-T$):

\begin{equation}
T_\theta=0.7236068 \label{eq 22theta}
\end{equation}
\begin{equation}
T_\phi=0.2763932 \label{eq 22phi1}
\end{equation}
\begin{equation}
 \mbox{Efficiency: } 0.2\times\frac{1}{5^2}
\end{equation}

This scheme generates a maximally entangled state of the form:
\[
\sim\frac{1}{\sqrt{5}}\Big[|40\>_{1',2'}+|31\>_{1',2'}+|22\>_{1',2'}+|13\>_{1',2'}+|04\>_{1',2'}\Big]
\]

The efficiency of producing Bell-type states falls exponentially as
the number input states is increased. We saw how increasing the
input from $|1\>_1|1\>_2$ to $|2\>_1|2\>_2$ led to an exponential
rise in the number of other possible outputs. Then, the probability
of obtaining the sub-selection (where all the detectors in
figure~\ref{epr(22)} register a single photon) from an
exponentially larger group of possible outcomes also falls
exponentially. This is apparent as we consider the
probabilities of obtaining Bell states from inputs $|3\>_1|3\>_2$
and $|4\>_1|4\>_2$.

\subsection*{Summary of results for beam-splitter transmittivities in EPR and BSA schemes}

\begin{table}[htbp]
\centering
\begin{tabular}{lcccccr}
\hline
Teleportee & Input & $T_\theta$ & $T_{\phi_1}$ & $T_{\phi_2}$ & $T_{\phi_3}$ & Efficiency\\
\hline $qubit$ & $|1\>|0\>$ & $0.5$ &  &  &  & $0.25$ (with PBS)\\
$qutrit$ & $|1\>|1\>$ & $0.211325$ &  &  &  & $0.06415$\\
$qupentit$ & $|2\>|2\>$ & $0.7236068$ & $0.2763932$ &  &  & $0.008$\\
$quheptit$ & $|3\>|3\>$ & $0.1510043$ & $0.6098260$ & $0.8495319$ &  & $5.306\times 10^{-5}$\\
$qunit$ & $|4\>|4\>$ & $0.2896110$ & $0.5212421$ & $0.8281260$ & $0.0399748$ & $5.4\times 10^{-9}$\\
\hline\label{tab transmitVals}
\end{tabular}
\caption{\ft{Summary of beam-splitter transmittivities required for even number inputs of photons 
to generate maximally entangled states for use in teleportation
schemes for qubits, qutrits, qupentits, quheptits, and qunits. This can be extended to odd number inputs for the teleportation of qutetrits,
quhexits, quoctits and beyond to arbitrary-dimentional states, or $|qudit\>$.}} \label{tab summaryTransmittivities}
\end{table}

\subsection*{A ``counting'' argument and the general scheme for the EPR source}
In the examples above for different symmetric photon inputs, a
general trend emerges that permits the formulation of a counting
argument for the linear optical EPR source we propose.
\\\\
Let us briefly return to the specific case of the photon input
$|1\>|1\>$ (section~\ref{secn input11}) required to produce a
Bell-state in a teleportation of a qutrit
($\alpha|0\>+\beta|1\>+\gamma|2\>$). In the relevant set-up on
figure~\ref{bs}~(ii), a Bell-state is generated by a single
beam-splitter whose transmittivity, $T$, is treated as a variable.
Thus we have one variable and one equation,~(\ref{eq epr11 1}),
which is solved for $T$ with the constraint ($T+R=1$)~(\ref{eq epr11
2}).
\\\\
The next input we considered, $|2\>|2\>$, was used to generate a
Bell-state containing five terms: $A|40\> + B|31\> + C|22\> \pm
B|13\> \pm A|04\>$. The amplitudes of the first two and the last two
terms are automatically equal by the symmetry of the scheme
(although there may be a phase difference). This leaves us with
three expressions for the first three amplitudes ($A,B,C$). Then, we
form two independent equations are formed, ($A=B$ and $B=C$) with
two variables.

We can see that for each additional pair of photons injected into
input $1$ and $2$ (on figure~\ref{epr(nn)}), there will be one
additional independent equation. 

\emph{Hence an arbitrary
symmetrical injection of $|n\>|n\>$ photons will give rise to $n+1$
amplitudes, from which it will be possible to form $n$ independent
equations. The number of independent variables, i.e.
transmittivities, in those equations is set by the number of
different beam splitters we introduce into the scheme. It follows that we require also $n$ independent transmittivities if we
are to solve $n$ equations. This requirement is ensured by
introducing $2n-1$ beam-splitters for an input of $|n\>|n\>$
photons. That is, each branch will consist of $n$ beam splitters,
the first of which is shared by the two branches.}
\\\\Figure~\ref{epr(nn)} shows the proposed
set-up for a generic EPR source for a symmetric input of number
state in its two branches, labelled $1$ and $2$.

A note on the scope of our argument: generally, the EPR scheme
produces Bell-type states for symmetric inputs of photons
($|n\>_{1}|n\>_{2}$). Maximally entangled states thus produced will
have the form expressed in equation~\ref{eq generalBell}:
\[
\sim \frac{1}{\sqrt{2n+1}}\Big[ |2n,0\>_{1',2'}+|(2n-1),1\>_{1',2'}+
\]
\begin{equation}
\ldots+|n,n\>_{1',2'}+\ldots+|1,(2n-1)\>_{1',2'}+|0,2n\>_{1',2'}\Big]
\label{eq generalBell}
\end{equation}

where n is the number of photons injected into each branch $(1$ and
$2)$ in figure~\ref{epr(nn)}.
\\
Such maximally entangled states can be used in teleportation of any
pure state with an \emph{odd number} of terms in the superposition:
that is, qutrits ($\alpha|0\>+\beta|1\>+\gamma|2\>)$, qupentits
($\alpha|0\>+\beta|1\>+\gamma|2\>+\mu|3\>+\lambda|4\>)$, quheptits
($\alpha|0\>+\beta|1\>+\gamma|2\>+\mu|3\>+\lambda|4\>+\xi|5\>+\zeta|6\>$)
and so on. Generating Bell-type states required to teleport
\emph{even number} arbitrary-dimensional pure states, such as
quadrits ($\alpha|0\>+\beta|1\>+\gamma|2\>+\mu|3\>$)and quhexits
($\alpha|0\>+\beta|1\>+\gamma|2\>+\mu|3\>+\lambda|4\>+\xi|5\>$),
requires an asymmetric input in the input channels $1$ and $2$.
However, our EPR scheme allows equally the generation of Bell-type
states with asymmetric inputs, which extends the range of teleportee
candidates to \emph{even number} superpositions too.
\\\\It has been suggested in \cite{calsamiglia,calsamiglia2,laflamme} that linear optical elements
would prove unamenable to the task of producing maximally entangled
states. Our findings suggest that their full potential remains as
yet unused and have much to contribute before resorting to non-linear
methods and the challenges associated with those.
\\\\The main limitation on the
higher orders remains the efficiency, which we see fall
exponentially for each step increase of the number of photons and
optical elements in the system. There is an additional drop in
efficiency due to the fact that only one of the Bell-states in any
given basis can be identified, going as $\frac{1}{d^2}$ for a
d-dimensional teleportee state.

\section*{The Bell State Analyser (BSA)}\label{secn bsa}

\begin{figure}
\includegraphics[width=15cm]{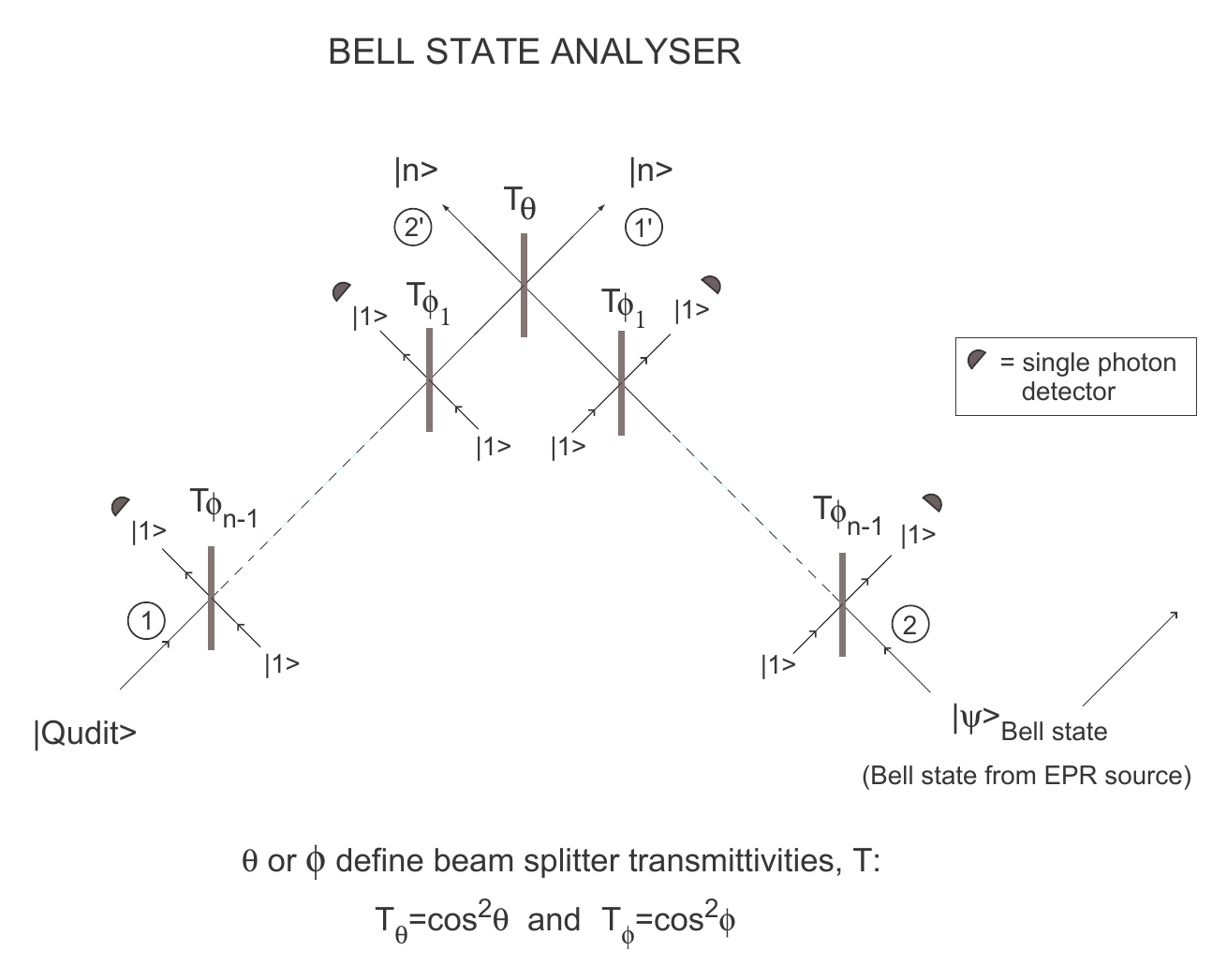}
\caption{Schematic of the Bell state analyser (BSA), reversing
all unitary operations performed by the EPR source. A Bell state can
be identified by the number of photons at the outputs labelled $1'$
and $2'$. For any given Bell state produced by our EPR source, the
outputs of the BSA would give back exactly the number of photons
used as inputs for the EPR source.}
\label{bsa(nn)}
\end{figure}

An advantage of the proposed the EPR source is that it can be
adapted for use as a Bell state analyser. This follows from the fact
that the scheme consists of a series of unitary operations carried
out by linear optical elements. If a Bell-state produced by our
scheme is then driven back through the system in time-reversal
fashion, that is, it undergoes in \emph{reverse sequence} the same
operations that produced it (starting with the last operation), we
can regain our input states. Thus by inverting our EPR scheme, as
shown in figure~\ref{bsa(nn)}, where the input is now a
Bell-state, the outputs are symmetrical numbers of photons
from which we can infer which Bell-state is present.

The BSA time-reversal measurement consists of applying the conjugate
unitary transformation for every optical element. This means that
what were previously the output states in the EPR scheme are now the
inputs for the BSA. Broadly, the scheme for generating the maximally
entangled states is reversed in the sense of undoing all the unitary
transformations in order to know what were its number-state inputs.
This is equivalent to knowing what Bell-state was present at the BSA
input.

\section*{Combining the EPR source and the BSA in teleportation}\label{secn combining EPR/bsa}

\begin{figure}
\includegraphics[width=15cm]{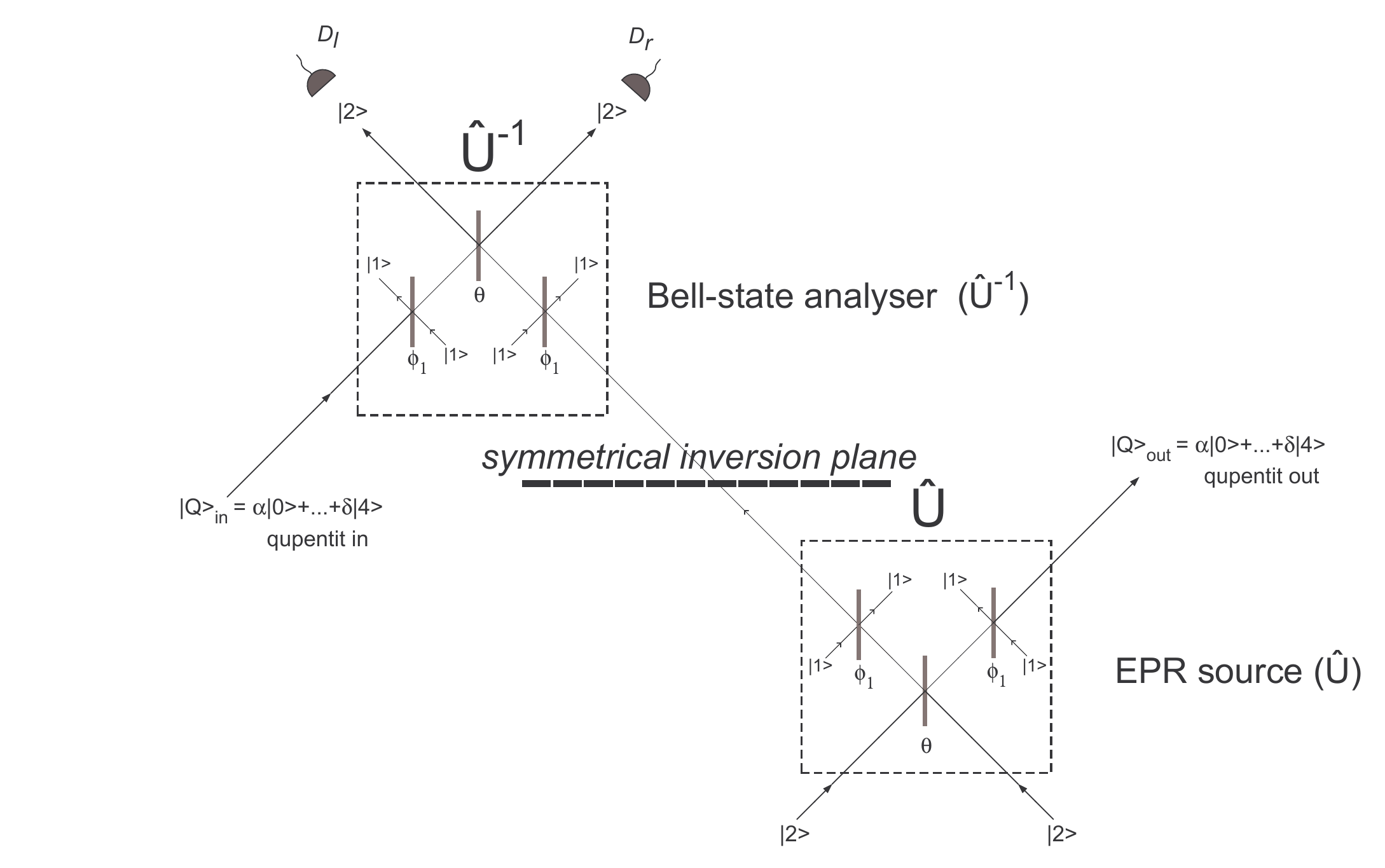}
\caption{Schematic showing how the two schemes for producing
maximally entangled states and then measuring come together in a
complete teleportation set-up. Since the EPR source consists of a
sequence of unitary operations $\mathbf{U}$, then to perform a
Bell-measurement, the BSA need only apply the conjugate
transformations in a time-reverse, $\mathbf{U}^\dagger$, fashion.
This makes the scheme symmetrical.}
\label{uu(11)}
\end{figure}

\begin{figure}
\includegraphics[width=\linewidth]{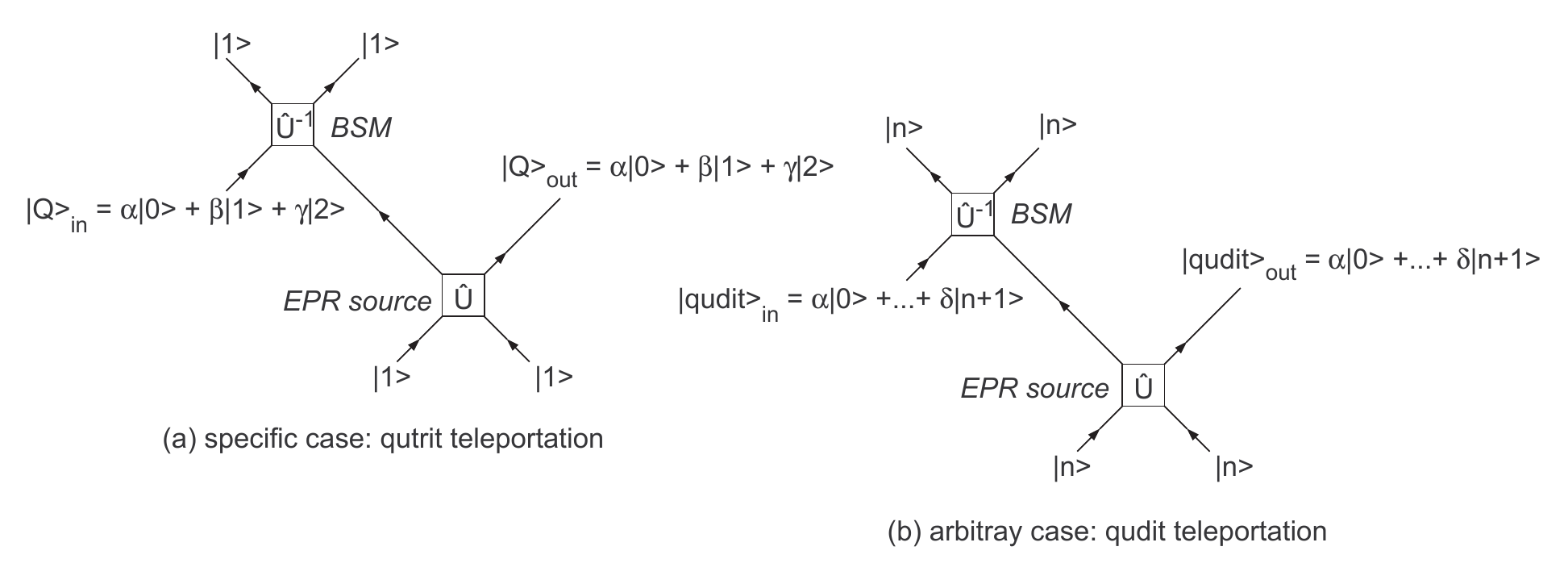}
\caption{Teleportation schematic for (a) a qutrit and (b) an
arbitrary-dimensional pure state, $|qudit\>$. The EPR source
ensures that the Bell-state required for the qudit is produced,
while the BSA applies the reverse operation to measure the Bell
state. As the dimension of the qudit increases, the efficiency of
teleportation falls exponentially.}
\label{uu}
\end{figure}

Figures~\ref{uu(11)} and~\ref{uu} show how the two schemes
for producing maximally entangled states and then measuring them
come together in a complete teleportation set-up.

In the example on figure~\ref{uu(11)}, the EPR source generates
by a series of unitary transformations $\mathbf{U}$, the maximally
entangled state required for the teleportation of a qutrit. This is
followed by a Bell-state analysis performed by the BSA, which
reverses the EPR source's operations by applying the conjugate
transformation $\mathbf{U}^\dagger$. Here, $\mathbf{U}$ and
$\mathbf{U^\dagger}$ represent all the individual operations
performed by each optical element in the scheme that produces and
measures the Bell state. This set-up has the property of being
symmetrical, which lends it more readily to implementation. Here,
the EPR source takes an input of $|1\>_1|1\>_2$ to generate one of
the Bell-type states required for a qutrit, while the BSA in this
example ``measures'' an output $|1\>|1\>$ with detectors $D_l$ and
$D_r$. This tells us that the Bell state into which the qutrit was
projected, must be the one produced by an input of $|1\>|1\>$ at the
EPR source.
\\A limitation of the scheme is that in its
simplest form, the BSA can identify only one of the possible nine
Bell states available to a qutrit. This entails a drop in the
teleportation efficiency to one-ninth of an ideal BSA. However,
using a polarising beam-splitter (PBS) can help to improve
efficiency, although this would still not remove the exponential
decay in efficiency for higher dimensional states.
\\\\Our teleportation scheme can be generalised to an
arbitrary-dimensional pure state, or $|qudit\>$ (figure~\ref{uu}(b))
since our EPR source can ensure a supply of maximally entangled
states of the correct form, while the BSA applies the reverse
operation. However, the teleportation efficiency is highly sensitive
to the dimension of the qudit, showing an exponential drop the
larger the qudit. This is a consequence of the exponential fall in
the efficiency of producing and detecting the Bell-states for
large qudit dimensions. The efficiency of the BSA falls also exponentially, since the measurement process is merely the reverse
of the production sequence. However, this problem can be avoided by keeping to relatively
low-dimensional qudits, and in particular, input states up to say $|2\>|2\>$
are well within experimental reach.

Another practical difficulty in the implementation of the teleportation scheme is the challenge of single-photon detection. One way of detecting single-photons using avalanche photodiodes or
photomultipliers, would be to introduce further cascades of
symmetric beam-splitters at the outputs. A ``click'' in a detector
does not tell us whether one, two or maybe three photons were
incident simultaneously. By introducing a symmetric beam-splitter
into a beam consisting of say two photons, we double the probability
that the ``click'' is caused by a single photon. Introducing yet
another beam-splitter would further increase the detection
reliability, albeit at the expense of the overall efficiency of the
scheme. However, this has become much less of a problem in recent years with the advent of highly efficient, low-noise, single-photon detectors.

\section*{Concluding remarks}
We began by introducing potential applications of number-state
teleportation as an effective method of information transfer in the
context of quantum information processing. We took as our starting
point a qubit teleportation protocol that we extended to an
arbitrary dimensional number states, using linear optical elements
exclusively. To this end, we presented a procedure for generating
maximally entangled states on demand and of the required dimension,
by using a cascade of beam-splitters arranged cross-beam fashion
along two axes in a symmetrical ``V'' layout. The key aspect of our
EPR scheme relied on treating the sequence of beam-splitter transmittivities along each axis as
variables. Their values were determined by the solutions to a set
simultaneous equations representing the amplitudes for the
sub-selection of number states forming the desired Bell-type state.
It was then shown how the scheme for the EPR source could easily be
adapted for use as a Bell-state analyser when implemented in a time-reversal manner. Finally, it was demonstrated that
when the output of the EPR scheme is used as the input to the BSA, we have a complete teleportation set-up for arbitrary
dimensional number-states.
\\\\Consideration was also given to the challenges associated with implementing this scheme. Teleportation efficiency decays exponentially for stepwise increases in the number-state dimension. This difficulty can be
avoided by using lower-dimensional states. In terms of applications,
it is unlikely that high-dimensional state teleportation would find
wide use, partly because of the technical difficulty of preparing
large pure states (i.e. the states to be teleported). However,
towards the lower-dimensional end, say, for qutrits, the
teleportation method proposed here maintains a high efficiency (see
section~\ref{tab transmitVals} for summary of values). It would would be worthwhile to see the scheme implemented experimentally as it would mark a practical milestone on the to quantum communication.

\section*{Acknowledgements}
The authors thank \v Caslav Brukner for discussions, and the Oxford Martin School and National Research Foundation Singapore for financial support.

\end{document}